\documentclass[epj]{svjour}
\usepackage{amsmath,amssymb}
\usepackage[utf8]{inputenc}
\usepackage{graphicx}% Include figure files
\usepackage{datetime}
\usepackage[pdftex,bookmarks,colorlinks]{hyperref}
\usepackage{tabularx}
\usepackage{multirow}
\usepackage{enumitem}
\usepackage[numbers,sort&compress]{natbib}
\newcommand\tabref[1]{\tablename~\ref{#1}}
\newcommand\secref[1]{sect.~\ref{#1}}
\newcommand\figref[1]{\figurename~\ref{#1}}
\begin{document}
\sloppy
\title{Restructuring of colloidal aggregates in shear flow}
\subtitle{Coupling interparticle contact models with Stokesian dynamics}

\author{Ryohei Seto\inst{1,2}
  \and Robert Botet\inst{3}
  \and Günter K. Auernhammer\inst{2}
  \and Heiko Briesen\inst{1}
}
\offprints{Ryohei Seto}
%(\email{setoryohei@me.com})
%\mail{R. Seto}
%
\institute{
  Chair for Process Systems Engineering, Technische Universität München, 
  Weihenstephaner Steig 23, 85350 Freising, Germany
  \and
  Max Planck Institute for Polymer Research,
  Ackermannweg 10, 55128 Mainz, Germany
  \and
  Laboratoire de Physique des Solides, UMR 8502, Université Paris-Sud, 
  91405 Orsay, France
}
\date{Received: date / Revised version: date}
\abstract{
  A method to couple interparticle contact models with Stokesian dynamics (SD)
  is introduced to simulate colloidal aggregates under flow conditions.
  The contact model mimics both the elastic and plastic behavior
  of the cohesive connections between particles within clusters.
  Owing to this, clusters can maintain their structures under low stress 
  while restructuring or even breakage may occur under sufficiently high stress conditions.
  SD is an efficient method to deal with the long-ranged and many-body nature of
  hydrodynamic interactions for low Reynolds number flows.
  By using such a coupled model, the restructuring of colloidal aggregates
  under shear flows with stepwise increasing shear rates was studied.
  Irreversible compaction occurs due to the increase of hydrodynamic stress on clusters.
  Results show that the greater part of the fractal clusters
  are compacted to rod-shaped packed structures,
  while the others show isotropic compaction.
}

\maketitle

\section{Introduction}

The mechanical properties of colloidal aggregates
are of fundamental interest in science and technology.
To classify particulate gels and to understand their rheological behaviors is a key element.
When attractive forces act among nano- or microscale particles,
they form finite-sized clusters or a space-filling network.
The latter shows a solid-like response to external stress,
so it is regarded as a gel.
In general, particulate gels are classified into two types
according to the attraction strength between particles~\cite{Larson_1999}.
If the attraction strength is sufficiently large,
the particle surfaces are deformed at the bonding point,
causing non-central forces~\cite{Johnson_1985}.
In this case, Brownian forces neither cause debonding nor tangential displacements
between contacting particles.
So, branched tenuous structures formed in the aggregation process 
are maintained~\cite{Lin_1989}.
On the other hand, if the attraction strength is weaker,
denser and multilinking local structures,
such as tetrahedral connections, are seen~\cite{Lu_2008}.
This can be explained by 
tangential displacements due to Brownian forces.
%

%%%%%%%%%%%%%%%%%%%%%%%%%%%%%%%%%%%%%%%%%%%%%%%%%%%%%%%%%%%%%%%%%%%%%%%%%%%%%%%%%%%%%%%%%%%%%%%%%%%%

The tangential displacements between contacting particles, 
\textit{i.e.} sliding, rolling and torsion,
play important roles in 
the structure formation and mechanical property of colloidal aggregates.
However, for nano or microscale particles,
it is not simple to characterize these interparticle interactions.
For example,
characterization of rolling resistances
requires elaborate experiments
such as AFM~\cite{Heim_1999} and optical tweezers~\cite{Pantina_2005}.
Though these direct observations have clearly proven
the existence of tangential forces,
the particles available for such measurements
are restricted to certain sizes.
This is why
there is still no general method
to fully characterize the contact forces in colloidal systems.
An alternative approach of investigating colloidal aggregates
is to develop simulation methods.
In particular, phenomena at the mesoscopic level
are expected to hold all necessary particle-scale information,
and the comparison between simulations and experimental observations
can be used for the characterization of contact forces.

%%%%%%%%%%%%%%%%%%%%%%%%%%%%%%%%%%%%%%%%%%%%%%%%%%%%%%%%%%%%%%%%%%%%%%%%%%%%%%%%%%%%%%%%%%%%%%%%%%%%

This work introduces a simulation method 
of coupling interparticle contact models and hydrodynamic interaction models.
The contact model used in this work
is similar to the one developed in granular physics%
~\citep{Iwashita_1998,Kadau_2002,Dominik_2002,Wada_2007,Luding_2008},
which is able to capture aggregates maintaining their structures under low stress
while being restructured under high stress.
The bond strength is assumed to be 
sufficiently larger than the thermal energy $k_{\mathrm{B}}T$,
therefore Brownian forces are not considered.
Instead, hydrodynamic stress induces restructuring of clusters.
The hydrodynamic interaction model employed here is Stokesian dynamics (SD)%
~\cite{Durlofsky_1987,Brady_1988,Ichiki_2002},
which provides the relations between velocities of particles
and the forces acting on them in the Stokes regime.
The evaluation of the hydrodynamic interactions
is the most difficult and time consuming part of the simulation
due to its long-ranged and many-body nature.
SD is based on Faxén’s law and 
multipole expansions to obtain the far-field mobility matrix,
which can simulate particle disturbed flows
with reasonable computational effort.

%%%%%%%%%%%%%%%%%%%%%%%%%%%%%%%%%%%%%%%%%%%%%%%%%%%%%%%%%%%%%%%%%%%%%%%%%%%%%%%%%%%%%%%%%%%%%%%%%%%%

We apply this simulation method 
to investigate the restructuring behavior of 
finite-sized tenuous clusters under flow conditions.
Investigation of colloidal aggregates under flow conditions
is a traditional problem in colloidal science. 
The original study on cluster sizes under shear flows
dates back to almost a century ago~\citep{Smoluchowski_1917},
which considered the cluster growth due to shear-induced collisions.
To estimate equilibrium cluster sizes,
one needs to know about the breakup mechanisms 
due to the hydrodynamic stress as well.
Theoretical studies of this problem appeared after decades~%
\cite{Bagster_1974,Adler_1979a,Sonntag_1986a},
and simulation studies of aggregate breakups
have been appearing over recent years~%
\cite{Potanin_1993,Higashitani_2001,Harada_2006,Becker_2008,
Becker_2009,Eggersdorfer_2010,Harshe_2011a}.
Restructuring of clusters is an additional 
and challenging issue in this context,
since it depends on details of the contact forces.
In order to focus on restructuring behavior,
a special situation is considered:
the shear flow is increased in a stepwise, thus less abrupt, 
manner than in previous works. 
In this case, the clusters are hardly broken; 
instead they are reinforced by new bonds 
generated during the restructuring process.
The time evolution of clusters
is expected to reflect the nature of contact forces.
Some characteristic restructuring behavior was observed 
in the following simulations.

%%%%%%%%%%%%%%%%%%%%%%%%%%%%%%%%%%%%%%%%%%%%%%%%%%%%%%%%%%%%%%%%%%%%%%%%%%%%%%%%%%%%%%%%%%%%%%%%%%%%

The contents of the paper are as follows:
the used methods, the contact model and SD,
are briefly described in \secref{sec_contact_model} and \secref{sec_method_SD}.
The coupling for the overdamped motion
is formulated in \secref{sec_overdamped_motion}. 
The optimization for the dilute limit of aggregate suspensions
is given in \secref{sec_optimization}.
Approaches to study the problem
are explained in \secref{sec_shearrate} and \secref{sec_stepwise_shear}.
After describing the parameters used for the simulations
in \secref{sec_parameters},
the results are shown by considering two main issues:
(i) how does the imposed shear flow result 
in the compaction of aggregates? (\secref{sec_compaction})
(ii) what is the tendencies of 
the shape formation and orientation? (\secref{sec_shape_orientation})
A discussion about the compaction 
in terms of consolidation is presented in \ref{sec_consolidation},
the observed tendencies in \ref{sec_reorientation},
and the hydrodynamic effect in \ref{sec_hydrodynamic_effect}.
Finally, the outcome of the work is concluded in \secref{sec_conclusion}.

 \section{Method}

\subsection{The contact model} 
\label{sec_contact_model}

\subsubsection{Model for the elasticity}
\begin{figure}[htb]
  \centering
  \includegraphics{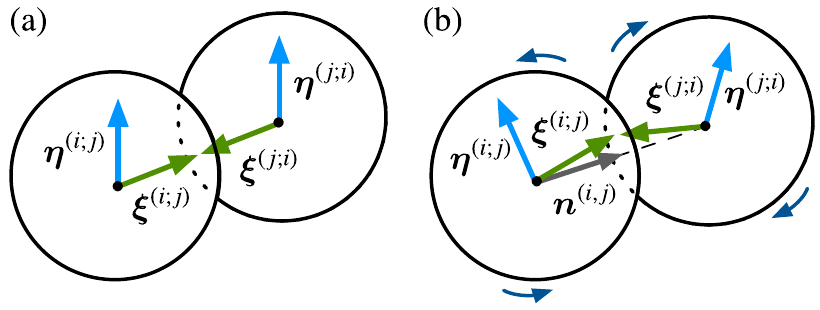}
  \caption{The contact-point indicators
    $\vec{\xi}^{(i;j)}$ and $\vec{\xi}^{(j;i)}$,
    and the torsion indicators
    $\vec{\eta}^{(i;j)}$ and $\vec{\eta}^{(j;i)}$,
    are illustrated for the stress-free state (a) and stressed state (b), respectively.
    The normal vector $\vec{n}^{(i,j)}$ always indicates the center-to-center direction.
  }
  \label{contact_modes}
\end{figure}

A simple contact model was employed
to simulate spherical particles cohesively connected.
The interaction between two particles 
is described by a cohesive bond
involving 4 types of degrees of freedom:
normal (the center-to-center direction),
sliding, bending%
\footnote{
We use `bending' instead of `rolling',
because they are equivalent except for a numerical factor,
but `bending' is more intuitive for dealing with deformations of colloidal aggregates.
}, and torsional displacements%
~\citep{Johnson_1985,sakaguchi_1993,Dominik_1997,Iwashita_1998,Zhang_1999,%
Kadau_2002,Dominik_2002,Delenne_2004,Jiang_2005,Tomas_2007,Wada_2007,%
Gilabert_2007,Luding_2008}.
These relative displacements
are expressed by using position vectors
and vectors fixed at respective particles.
So, a rotation of the frame of reference 
does not affect the result, i.e. objectivity is satified~\cite{Luding_2008}.
In this work, 
the Hookean force-displacement relationships
are assumed for these degrees of freedom,
which are characterized by the spring constants
$k_{\mathrm{N}}$, $k_{\mathrm{S}}$, $k_{\mathrm{B}}$, and $k_{\mathrm{T}}$.
\begin{description}[leftmargin=0pt]
\item[\emph{Normal displacement}]
Let us suppose two spherical particles $i$ and $j$
located at $\vec{r}^{(i)}$ and $\vec{r}^{(j)}$.
The center-to-center distance $r^{(i,j)} \equiv |\vec{r}^{(i)} - \vec{r}^{(j)}|$
is changed by the normal element of the acting force.
A monodisperse system is considered here,
so the radius of particle is denoted by $a$.
The force-displacement relation is given by
\begin{equation}
\vec{F}_{\mathrm{N}}^{(i,j)} =  k_{\mathrm{N}} (r^{(i,j)} - 2a) \, \vec{n}^{(i,j)},
\end{equation}
where $\vec{n}^{(i,j)} \equiv (\vec{r}^{(j)} - \vec{r}^{(i)}) /r^{(i,j)} $
is the normal direction.
\item[\emph{Sliding displacement}]
Sliding displacement is a tangential element
of the relative displacement between particles with fixed orientations.
In order to express the sliding displacement vector $\vec{d}^{(i,j)}$,
unit vectors fixed to each particle:
$\vec{\xi}^{(i;j)}$ and $\vec{\xi}^{(j;i)}$, were introduced,
called \emph{contact-point indicators} in this paper (\figref{contact_modes}).
Using the indicators, the positions of the original contact points are written 
as follows:
\begin{equation}
\vec{r}_{\mathrm{o.c.}}^{(i;j)} = \vec{r}^{(i)} + a \, \vec{\xi}^{(i;j)}, \quad
\vec{r}_{\mathrm{o.c.}}^{(j;i)} = \vec{r}^{(j)} + a \, \vec{\xi}^{(j;i)}.
\end{equation}
When two particles get in contact, \textit{i.e.} at the stress-free state,
the contact points are the same
$\vec{r}_{\mathrm{o.c.}}^{(i;j)} = \vec{r}_{\mathrm{o.c.}}^{(j;i)} $,
and the contact-point indicators are set to
$\vec{\xi}^{(i;j)} = \vec{n}^{(i,j)}$ [(a) in \figref{contact_modes}].
The sliding displacement vector is given by 
the projection of the deviation
$\vec{r}_{\mathrm{o.c.}}^{(j;i)} - \vec{r}_{\mathrm{o.c.}}^{(i;j)} $
onto the perpendicular bisector between the two particles:
\begin{align}
\vec{d}^{(i,j)}
&\equiv
\vec{r}_{\mathrm{o.c.}}^{(j;i)} - \vec{r}_{\mathrm{o.c.}}^{(i;j)} 
- \bigl\{(\vec{r}_{\mathrm{o.c.}}^{(j;i)} - \vec{r}_{\mathrm{o.c.}}^{(i;j)} 
)\cdot  \vec{n}^{(i,j)} \bigr\}\vec{n}^{(i,j)} \notag 
\\& = 
a \bigl\{
\Delta \vec{\xi}^{(i,j)}
- ( \Delta \vec{\xi}^{(i,j)} \cdot \vec{n}^{(i,j)} ) \vec{n}^{(i,j)}
\bigr\},
\end{align}
where $\Delta \vec{\xi}^{(i,j)} \equiv \vec{\xi}^{(j;i)} - \vec{\xi}^{(i;j)}$.
So, the force-displacement relation is given by
\begin{equation}
 \vec{F}_{\mathrm{S}}^{(i,j)}  = k_{\mathrm{S}} \vec{d}^{(i,j)}.
\end{equation}
\item[\emph{Bending displacement}]
Bending is a type of tangential displacement involving rotation,
with the angle between the contact-point indicators quantifying this displacement.
This angle is assumed to be small,
so it can be approximated by the norm of the vector product
$| \vec{\xi}^{(j;i)} \times ( - \vec{\xi}^{(i;j)}) |$.
Since it includes some torsional element,
the bending angle vector $\vec{\varphi}^{(i,j)} $
is obtained by subtracting the normal part:
\begin{equation}
  \vec{\varphi}^{(i,j)} \equiv
  - \vec{\xi}^{(j;i)} \times \vec{\xi}^{(i;j)} 
  +
  \bigl\{( \vec{\xi}^{(j;i)} \times  \vec{\xi}^{(i;j)} ) \cdot \vec{n}^{(i,j)}
    \bigr\}\vec{n}^{(i,j)}.
\end{equation}
By using the bending angle vector, 
the moment-angle relation is given by
\begin{equation}
  \vec{M}^{(i,j)}_{\mathrm{B}} =
  k_{\mathrm{B}} a^2 \vec{\varphi}^{(i,j)}.
  \label{moment_bending}
\end{equation}
\item[\emph{Torsional displacement}]
Torsion is the rotational displacement around the normal vector $\vec{n}^{(i;j)}$.
In order to express the torsional angle,
another set of unit vectors fixed to each particle:
$\vec{\eta}^{(i;j)}$ and $\vec{\eta}^{(j;i)}$,
are introduced, 
called \emph{torsion indicators} (\figref{contact_modes}).
When two particles get in contact, \textit{i.e.} at the stress-free state,
they are set by choosing ones from the vectors being 
orthogonal to the normal vector:
$\vec{\eta}^{(i;j)} \cdot \vec{n}^{(i;j)} = 0$
and $\vec{\eta}^{(j;i)} \cdot \vec{n}^{(i;j)} =0$,
and parallel to each other $\vec{\eta}^{(i;j)} = \vec{\eta}^{(j;i)}$
[(a) in \figref{contact_modes}].
Since the torsional angle is also assumed to be small,
it can be approximated by the norm of the vector product 
$|\vec{\eta}^{(i;j)} \times \vec{\eta}^{(j;i)}|$.
The torsional angle vector $\vec{\theta}^{(i,j)}$ 
is defined as the normal element of their vector product:
\begin{equation}
  \vec{\theta}^{(i,j)}
\equiv
\bigl\{(\vec{\eta}^{(i;j)} \times \vec{\eta}^{(j;i)}) \cdot \vec{n}^{(i,j)}\bigr\}
\vec{n}^{(i,j)}.
\end{equation}
By using the torsional angle vector,
the moment-angle relation is given by
\begin{equation}
\vec{M}_{\mathrm{T}}^{(i,j)} = k_{\mathrm{T}} a^2 \vec{\theta}^{(i,j)}.
\end{equation}
\end{description}

%%%%%%%%%%%%%%%%%%%%%%%%%%%%%%%%%%%%%%%%%%%%%%%%%%%%%%%%%%%%%%%%%%%%%%%%%%%%%%%%%%%%%%%%%%%%%%%%%%%%
Thus, the forces and moments on the contact point between
particle $i$ and $j$ are related to the corresponding displacements.
The force and torque acting on the particle $i$
from the contacting particles $j$ are given by their sums:
\begin{equation}
  \begin{split}
    \vec{F}_{\mathrm{P}}^{(i)} &=
    \sum_j 
    \left(
      \vec{F}_{\mathrm{N}}^{(i,j)}
      + \vec{F}_{\mathrm{S}}^{(i,j)}
    \right) 
    , \\
    \vec{T}_{\mathrm{P}}^{(i)} &= 
    \sum_j
    \left(
      a \, \vec{n}^{(i,j)} \times   \vec{F}_{\mathrm{S}}^{(i,j)}
      + \vec{M}_{\mathrm{B}}^{(i,j)} 
      + \vec{M}_{\mathrm{T}}^{(i,j)}
    \right)
    .
  \end{split}
  \label{contact_force_and_torque}
\end{equation}
The suffix P indicates the particle contact interactions
in contrast to the hydrodynamic interactions.

\subsubsection{Model for the plasticity}
\label{method_plasticity}

In the contact model,
the potential energy is stored in the introduced bonds
as long as the stresses acting on the bonds are small.
When stresses become larger than a certain threshold,
the bond breaks and the stored energy is dissipated.
If particles are still in contact,
the contact-point indicators and torsion indicators are reset with the configuration
to release the potential energy stored in the tangential springs.

%%%%%%%%%%%%%%%%%%%%%%%%%%%%%%%%%%%%%%%%%%%%%%%%%%%%%%%%%%%%%%%%%%%%%%%%%%%%%%%%%%%%%%%%%%%%%%%%%%%%

The supportable strength for a cohesive bond depends on 
the direction of the acting forces or moments.
So, the breakableness is characterized by 
two critical forces and two critical moments:
$F_{\mathrm{Nc}}$, $F_{\mathrm{Sc}}$, $M_{\mathrm{Bc}}$, and $M_{\mathrm{Tc}}$.
In general, all components of the bond are stressed simultaneously.
This is why a criterion of breakage 
can be introduced by a destruction function
$\zeta(F_{\mathrm{N}}, F_{\mathrm{S}},M_{\mathrm{B}}, M_{\mathrm{T}})$,
whose positive value indicates breakage.
Here, a simple energy-like function is used~\citep{Delenne_2004}:
\begin{equation}
 \zeta 
= 
\vartheta(F_{\mathrm{N}})
\frac{F_{\mathrm{N}}^2}{F_{\mathrm{Nc}}^2}
+
\frac{F_{\mathrm{S}}^2}{F_{\mathrm{Sc}}^2}
+
\frac{M_{\mathrm{B}}^2}{M_{\mathrm{Bc}}^2}
+
\frac{M_{\mathrm{T}}^2}{M_{\mathrm{Tc}}^2}
 - 1,
\label{destruction_function}
\end{equation}
where $\vartheta(F_{\mathrm{N}})$ is Heaviside function
$\vartheta(F_{\mathrm{N}})= 1$ for $F_{\mathrm{N}} \geq 0$ 
and $\vartheta(F_{\mathrm{N}})= 0$ for $F_{\mathrm{N}} < 0$.

%%%%%%%%%%%%%%%%%%%%%%%%%%%%%%%%%%%%%%%%%%%%%%%%%%%%%%%%%%%%%%%%%%%%%%%%%%%%%%%%%%%%%%%%%%%%%%%%%%%%

According to the intensive studies by \citet{Dominik_1997},
the critical normal and sliding forces,
$F_{\mathrm{Nc}}$ and $F_{\mathrm{Sc}}$,
are much larger than the corresponding forces
of the critical bending and torsional moments,
$M_{\mathrm{Bc}}/a$ and $M_{\mathrm{Tc}}/a$.
For a typical case, the ratio can be the order of $10^2$.
Thus, this work focuses on the effects for the bending 
and torsional breakups
and excludes the separation and sliding breakups.
Besides,
the direct measurements of the critical bending moment
have been reported \cite{Heim_1999,Pantina_2005},
while no direct measurement is available for the critical torsional moment.
For simplicity the same strength for the bending 
and torsional moments is assumed here.
In short,
a special case of the bond breakableness written by
$F_{\mathrm{Nc}}\to \infty$ and $F_{\mathrm{Sc}}\to \infty$
and $M_{\mathrm{Bc}} = M_{\mathrm{Tc}} = M_{\mathrm{c}}$
is considered.
The strength of bond is then given with one parameter $M_{\mathrm{c}}$ by 
\begin{equation}
  \zeta = 
  \frac{ M_{\mathrm{B}}^2 + M_{\mathrm{T}}^2 }{M_{\mathrm{c}}^2} - 1.
  \label{simplified_destruction_functions}
\end{equation}

\subsubsection{Model for the new connection}
\label{method_new_bond}

As of now, 
interactions between contacting particles are defined,
but no assumption about particles being
initially farther apart has been made.
We consider a short-range cohesive interaction in this work.
As a simple case, it is assumed that 
no interaction acts between remote particles ($r^{(i,j)} > 2a$).
If, however, two particles approach each other and get into contact,
\textit{i.e.} $r^{(i,j)}= 2a$,
they start to interact with each other,
which is modeled by the generation of a cohesive bond.

\subsection{Hydrodynamic interaction (Stokesian dynamics)}
\label{sec_method_SD}

Stokesian dynamics (SD) is employed
for evaluating the hydrodynamic interactions~%
\cite{Durlofsky_1987,Brady_1988,Ichiki_2002}.
Here,
simple shear flows $\vec{u}^{\infty}(\vec{r}) = z \dot{\gamma} \vec{e}_{x}$ 
are considered to apply, where $\dot{\gamma}$ is the shear rate.
The force-torque-stresslet (FTS) version of SD 
is required to solve the flow conditions.
By using the translational velocity $\vec{U}^{\infty}$,
vorticity $\vec{\Omega}^{\infty}$, and rate-of-strain $\tens{E}^{\infty}$,
the flow field $\vec{u}^{\infty} (\vec{r})$ is expressed as follows:
\begin{equation}
  \vec{u}^{\infty} (\vec{r})
  = \vec{U}^{\infty} + \vec{\Omega}^{\infty} \times \vec{r}
  + \tens{E}^{\infty} \vec{r}.
  \label{equation_linear_flows}
\end{equation}
with the following nonzero elements:
$\Omega^{\infty}_y = \dot{\gamma}/2$
and
$E^{\infty}_{xz} = E^{\infty}_{zx}= \dot{\gamma}/2$.
The hydrodynamic interactions acting on a particle $i$,
\textit{i.e.} the drag force $\vec{F}^{(i)}_{\mathrm{H}}$, torque $\vec{T}^{(i)}_{\mathrm{H}}$, 
and stresslet $\vec{S}^{(i)}_{\mathrm{H}}$, 
are given as linear combinations of the relative velocities from the imposed flow:
the translational and rotational velocities
$\vec{U}^{(j)} - \vec{u}^{\infty}(\vec{r}^{(j)})$
and $\vec{\Omega}^{(j)}-\vec{\Omega}^{\infty}$,
of all particles ($j = 1, \dotsc,N$) and the rate of strain $-\vec{E}^{\infty}$.
The linear combinations for all particles
are expressed as a matrix form
\begin{equation}
  \begin{pmatrix}
    \vec{F}_{\mathrm{H}} \\
    \vec{T}_{\mathrm{H}} \\
    \vec{S}_{\mathrm{H}} 
  \end{pmatrix}
  = -
  \tens{R}
  \begin{pmatrix}
    \vec{U} - \vec{U}^{\infty}(\vec{r}) \\
    \vec{\Omega} - \vec{\Omega}^{\infty} \\
    - \vec{E}^{\infty}
  \end{pmatrix},
  \label{resistance_form}
\end{equation}
where the vectors 
involve $11 N$ elements for all particles,
and the matrix $\tens{R}$
is the so-called grand resistance matrix%
\footnote{
Since both the stresslet
and rate-of-strain tensors are symmetric and traceless, the five
independent elements are denoted as vector forms, such as
$\vec{S} \equiv (S_{xx},S_{xy},S_{xz},S_{yz},S_{yy})$.
}.
%

%%%%%%%%%%%%%%%%%%%%%%%%%%%%%%%%%%%%%%%%%%%%%%%%%%%%%%%%%%%%%%%%%%%%%%%%%%%%%%%%%%%%%%%%%%%%%%%%%%%%

It must be noted that
the lubrication correction of SD is not applied in this work.
For suspensions where the interparticle interaction is absent,
the lubrication forces play essential role 
for near contact particles~\citep{Phung_1996,Foss_2000}.
On the other hand, 
for rigid clusters, 
i.e. if the relative velocities between particles are zero 
due to strong cohesive forces,
the lubrication correction has no contribution.
Thus, the lubrication correction to the mobility matrix can 
safely be omitted~\cite{Bossis_1991,Harshe_2010,Seto_2011}.
Though, the relative velocities between particles are not perfectly zero in this work,
near rigid-motion of clusters is investigated 
by only gradually increasing the shear rate. 
Due to the resulting small relative velocities of the primary particles, 
the lubrication correction is expected to be less important.
The neglect of lubrication forces is also a necessity for 
the computational approach presented later. 
The investigated simulation times are only accessible with reasonable computational effort, 
if the time-scales of the long-range hydrodynamic forces 
and short-range contact forces can be separated.
This separability allows the reuse of the mobility matrix 
for several time steps which significantly enhances computational performance.
This would unfortunately not be the case if lubrication forces would be considered.

\subsection{Overdamped motion}
\label{sec_overdamped_motion}
%{\bf check to read \cite{Ball_1997}}

%
To simulate the time evolution of particles with contact models,
configurations of spherical particles are
described by not only their central positions $\vec{r}^{(i)}(t)$ ($i=1,\dotsc,N$),
but also by their orientations.
The orientation of a particle $i$ is expressed by 
using a quaternion $\tilde{q}^{(i)}(t)$.
If we set $\tilde{q}^{(i)}(0)=1$ at the initial time,
the quaternion $\tilde{q}^{(i)}(t)$ means the rotation 
from the initial orientation.
The rotation of a vector $\vec{\xi}$ fixed to the particle
is written as 
$\vec{\xi}(t) = \tilde{q}^{(i)}(t) \vec{\xi}(0) \{\tilde{q}^{(i)}(t) \}^{-1} $.

%%%%%%%%%%%%%%%%%%%%%%%%%%%%%%%%%%%%%%%%%%%%%%%%%%%%%%%%%%%%%%%%%%%%%%%%%%%%%%%%%%%%%%%%%%%%%%%%%%%%

When the contact forces are strong enough,
Brownian forces are negligible.
But, the hydrodynamic forces depend on the imposed shear flow.
Therefore, 
only contact and hydrodynamic forces acting on the particles were considered.
In general, the particles follow the Newton's equations of motion:
\begin{equation}
  m \frac{d \vec{U}}{dt}
  = \vec{F}_{\mathrm{P}} + \vec{F}_{\mathrm{H}}, \quad 
  I \frac{d \vec{\Omega}}{dt}
  = \vec{T}_{\mathrm{P}} + \vec{T}_{\mathrm{H}},
  \label{full_eq_of_motion}
\end{equation}
where $m$ and $I$ are the mass and moment of inertia of the particles, respectively.
The velocities $\vec{U}$, angular velocities $\vec{\Omega}$, 
forces $\vec{F}$ and torques $\vec{T}$ include $N$ vectors
for all particles.
%

%%%%%%%%%%%%%%%%%%%%%%%%%%%%%%%%%%%%%%%%%%%%%%%%%%%%%%%%%%%%%%%%%%%%%%%%%%%%%%%%%%%%%%%%%%%%%%%%%%%%

For colloidal systems,
the inertia of particles are negligibly small
compared to the hydrodynamic forces.
By neglecting the inertia terms,
the equations of motion \eqref{full_eq_of_motion}
are approximated by the force- and torque-balance equations:
\begin{equation}
    \vec{F}_{\mathrm{P}} + \vec{F}_{\mathrm{H}} \approx 0,
    \quad
    \vec{T}_{\mathrm{P}} + \vec{T}_{\mathrm{H}} 
    \approx 0.
  \label{balance_equations}
\end{equation}
Systems following these balance equations are called overdamped.
In order to solve the overdamped motion with SD,
the mobility form:
\begin{equation}
  \begin{pmatrix}
    \vec{U} - \vec{U}^{\infty} \\
    \vec{\Omega} - \vec{\Omega}^{\infty} \\
    \vec{S}_{\mathrm{H}}
  \end{pmatrix}
  = 
-
  \tens{M}
  \begin{pmatrix}
    \vec{F}_{\mathrm{H}} \\
    \vec{T}_{\mathrm{H}} \\
    - \vec{E}^{\infty}
  \end{pmatrix},
  \label{mobility_form}
\end{equation}
is used instead of the resistance form \eqref{resistance_form}.
The numerical library developed by Ichiki~\citep{Ichiki_2006}
was used to obtain the mobility matrix $\tens{M}$.
By combining \eqref{balance_equations} and
\eqref{mobility_form},
the velocities of the particles $(\vec{U}, \vec{\Omega})$ 
are given by the functions of contact interactions 
$(\vec{F}_{\mathrm{P}}, \vec{T}_{\mathrm{P}})$:
\begin{equation}
  \vec{U}(t) = 
  \vec{U}( 
    \vec{F}_{\mathrm{P}}, 
    \vec{T}_{\mathrm{P}}), \quad 
  \vec{\Omega}(t) = 
  \vec{\Omega}( 
    \vec{F}_{\mathrm{P}}, 
    \vec{T}_{\mathrm{P}}).
\end{equation}
Once their velocities are determined, 
the time evolution of the particles
are given by integrating the time derivative relations:
\begin{equation}
  \begin{split}
    \frac{\mathrm{d} \vec{r}^{(i)}}{\mathrm{d}t} &= \vec{U}^{(i)}, \quad
    %%%%%%%%%%%%
    \frac{\mathrm{d} \tilde{q}^{(i)}}{\mathrm{d}t} 
    = 
    \hat{\tens{\Omega}}^{(i)} 
    \tilde{q}^{(i)},
  \end{split}
  \label{eqs_time_derivative_relations}
\end{equation}
where the matrix $\hat{\mathsf{\Omega}}^{(i)} $ is constructed 
by the elements of the angular velocity $\vec{\Omega}^{(i)}$
as follows:
\begin{equation}
  \hat{\tens{\Omega}}^{(i)}  \equiv
 \begin{pmatrix}
    0       & -\Omega_x^{(i)} & -\Omega_y^{(i)} & -\Omega_z^{(i)} \\
    \Omega_x^{(i)} & 0               & -\Omega_z^{(i)} &  \Omega_y^{(i)} \\
    \Omega_y^{(i)} &  \Omega_z^{(i)} & 0               & -\Omega_x^{(i)} \\
    \Omega_z^{(i)} & -\Omega_y^{(i)} & \Omega_x^{(i)}  & 0
  \end{pmatrix}.
\end{equation}
Since the overdamped motions with the simplified contact model 
and approximated hydrodynamics are considered,
the accuracy of the numerical integration has no primary importance.
Therefore, the explicit Euler method 
was used to integrate the differential equations \eqref{eqs_time_derivative_relations}
with a discretized time step $\delta t$.

%%%%%%%%%%%%%%%%%%%%%%%%%%%%%%%%%%%%%%%%%%%%%%%%%%%%%%%%%%%%%%%%%%%%%%%%%%%%%%%%%%%%%%%%%%%%%%%%%%%%

\subsection{Reusing the mobility matrix for deforming clusters}
\label{sec_optimization}
The bottle neck to simulate the time evolution
is the calculation of 
the mobility matrix $\tens{M}$ in \eqref{mobility_form} in each time step.
Since the contact forces are changed by short displacements of particles,
the time step $\delta t$ needs to be set small enough,
causing a large calculational effort.
This is why a way to reduce the computational effort 
has to be introduced.

%%%%%%%%%%%%%%%%%%%%%%%%%%%%%%%%%%%%%%%%%%%%%%%%%%%%%%%%%%%%%%%%%%%%%%%%%%%%%%%%%%%%%%%%%%%%%%%%%%%%

The mobility matrix $\tens{M}$ depends only on the positions of particles.
If the relative positions of particles within an isolated cluster remain unchanged,
the hydrodynamic interactions 
under any flow written as in \eqref{equation_linear_flows}
can be evaluated with a single mobility matrix.
Though clusters are not rigid in this work,
up to a certain degree the deformed structure can 
be considered the same for the hydrodynamic interactions.
As long as the deformation is negligible in this sense,
a mobility matrix may be reused repeatedly.
%

%%%%%%%%%%%%%%%%%%%%%%%%%%%%%%%%%%%%%%%%%%%%%%%%%%%%%%%%%%%%%%%%%%%%%%%%%%%%%%%%%%%%%%%%%%%%%%%%%%%%

In order to evaluate the motion of an isolated cluster,
one can take the center-of-mass of a cluster
as the origin of the coordinate without loss of generality.
Let us suppose that the deformation of the structure of the cluster 
for a time interval $\Delta t$ is negligible.
In this case,
the time evolution of the particles
from $t$ to $t' = t + \Delta t$
can be approximated by 
\begin{equation}
  \vec{r}^{(i)} (t') \approx \tens{R}_{t\to t'}\vec{r}^{(i)} (t),
\end{equation}
where $\tens{R}_{t\to t'}$ is a rotation matrix.
The hydrodynamic interaction at the time $t'$,
\textit{i.e.} the relations between
$(\vec{F}^{(i)}_{\mathrm{H}}(t') , \vec{T}^{(i)}_{\mathrm{H}}(t') )$
and $ (\vec{U}^{(i)}(t'),   \vec{\Omega}^{(i)}(t') )$
can be obtained by using the mobility matrix at the time $t$ as follows:
\begin{equation}
  \begin{pmatrix}
    \Delta \bar{\vec{U}} (t') \\
    \Delta \bar{\vec{\Omega} }(t') \\
    \bar{\vec{S}}_{\mathrm{H}} (t')
  \end{pmatrix}
  =
-
  \tens{M}(t)
  \begin{pmatrix}
    \bar{\vec{F}}_{\mathrm{H}} (t')\\
    \bar{\vec{T}}_{\mathrm{H}} (t')\\
    - \bar{\vec{E}}^{\infty} (t')
  \end{pmatrix}
\end{equation}
where one has the following relations:
\begin{equation}
  \begin{split}
    \bar{\vec{F}}^{(i)}_{\mathrm{H}} (t')
    &= 
    \tens{R}^{-1}_{t \to t'}\vec{F}^{(i)}_{\mathrm{H}}(t') , \\
    %%%%%%%%%%%%
    \bar{\vec{T}}^{(i)}_{\mathrm{H}}(t')
    &= 
    \tens{R}^{-1}_{t \to t'}  \vec{T}^{(i)}_{\mathrm{H}}(t') , \\
    %%%%%%%%%%%%
    \bar{\tens{E}}^{\infty} (t')
    &=
    \tens{R}^{-1}_{t \to t'} \tens{E}^{\infty} 
    \tens{R}_{t \to t'},
  \end{split}
\end{equation}
and
\begin{equation}
  \begin{split}
    \vec{U}^{(i)}(t') &= 
    \tens{R}_{t \to t'}  \Delta \bar{\vec{U}}^{(i)} + \vec{U}^{\infty}(\vec{r}^{(i)}(t')),\\
    %%%%%%%%
    \vec{\Omega}^{(i)}(t') &= 
    \tens{R}_{t \to t'} \Delta \bar{\vec{\Omega}}^{(i)} + \vec{\Omega}^{\infty}. 
  \end{split}
\end{equation}
Now one needs to determine the rotation matrix $\tens{R}_{t \to t'}$
for the cluster, 
which is deformed during the actual time evolution.
For a trial rotation matrix $\tens{R}$,
the positions of particles $\vec{r}^{(i)}(t)$ are transformed to
\begin{equation}
\vec{s}^{(i)}  = \tens{R} \, \vec{r}^{(i)}(t).
\end{equation}
The optimal rotation matrix $\tens{R}_{\mathrm{opt}}$ should minimize
the differences between the actual positions $\vec{r}^{(i)}(t')$
and the transformed positions $\vec{s}^{(i)}$.
One can take the following objective function to be minimized:
\begin{equation}
 D(\tens{R})
 \equiv
\frac{1}{N}
\sum_{i} \bigl\{\vec{r}^{(i)}(t') - \vec{s}^{(i)} \bigr\}^2
\end{equation}
The gradient descent method is employed to find 
the optimal rotation matrix $\tens{R}_{\mathrm{opt}}$.
Thus, 
the rotation matrix $\tens{R}_{t \to t'}$
can be determined: $\tens{R}_{t \to t'} = \tens{R}_{\mathrm{opt}}$.

%%%%%%%%%%%%%%%%%%%%%%%%%%%%%%%%%%%%%%%%%%%%%%%%%%%%%%%%%%%%%%%%%%%%%%%%%%%%%%%%%%%%%%%%%%%%%%%%%%%%

The objective function 
with the optimal rotation $D(\tens{R}_{t \to t'})$
represents the degree of the deformation.
If the deformation of the cluster becomes larger than 
a threshold:
$D(\tens{R}_{t \to t'}) \geq D_{\mathrm{max}}$,
the mobility matrix needs to be updated.

\subsection{Introduction of a dimensionless shear rate}
\label{sec_shearrate}

The behavior of a cluster formed by strong cohesion under a strong flow 
is equivalent to the case of weak cohesion and a weak flow.
In order to reduce the redundancy,
a dimensionless variable, the ratio between hydrodynamic interactions and contact forces,
is introduced.
The cohesive force gives the typical force of the simulation $F_{0}$;
the critical force for bending and torsional breakages 
is taken for that: $F_{0} = M_{\mathrm{c}}/a$,
because they play an important role for the restructuring of tenuous clusters.
Since hydrodynamic interactions
are proportional to the shear rate $\dot{\gamma}$ in the Stokes regime,
the dimensionless shear rate can be defined by
$ \dot{\Gamma} \equiv 6 \pi \eta_0 a^2 \dot{\gamma}/ F_{0} 
= 6 \pi \eta_0 a^3 \dot{\gamma} / M_{\mathrm{c}}$,
which indicates the flow strength for the contact force.
In this work,
the shear-rate dependence
is discussed in terms of this dimensionless variable $\dot{\Gamma}$.

\subsection{Stepwise increase of shear rates}
\label{sec_stepwise_shear}

In general,
three types of behaviors are expected 
for a cluster in shear flows:
\begin{description}[leftmargin=0pt]
  \item[\emph{Rigid body rotation}]
  When the hydrodynamic stress is sufficiently weak,
  the cluster rotates without changing its structure.
  \item[\emph{Restructuring}]
  When the hydrodynamic stress slightly exceeds the strength of the cluster,
  the cluster is restructured.
  Newly generated cohesive bonds during the restructuring
  may reinforce the cluster.
  If the strength of the cluster exceeds the hydrodynamic stress,
  it turns to the `rigid body rotation' regime.
  \item[\emph{Breakup}]
  When the hydrodynamic stress is much stronger than the strength of the cluster,
  the cluster is significantly elongated and may be broken up into smaller pieces.
\end{description}
In other simulation studies
\cite{Potanin_1993,Higashitani_2001,Harada_2006,
  Zeidan_2007,Becker_2009,Becker_2010,Eggersdorfer_2010,Harshe_2011a},
the shear flow is abruptly applied as a step function of time.
In that case,
the restructuring plays a limited role in
a certain range of shear rates.
This  change of shear rate in a single step is a simple
but very special case in terms of shear history.
If the flow strength is less abruptly increased,
restructuring may reinforce the cluster before reaching higher shear rates.
%
%%%%%%%%%%%%%%%%%%%%%%%%%%%%%%%%%%%%%%%%%%%%%%%%%%%%%%%%%%%%%%%%%%%%%%%%%%%%%%%%%%%%%%%%%%%%%%%%%%%%

\begin{figure}[htb]
  \centering
  \includegraphics{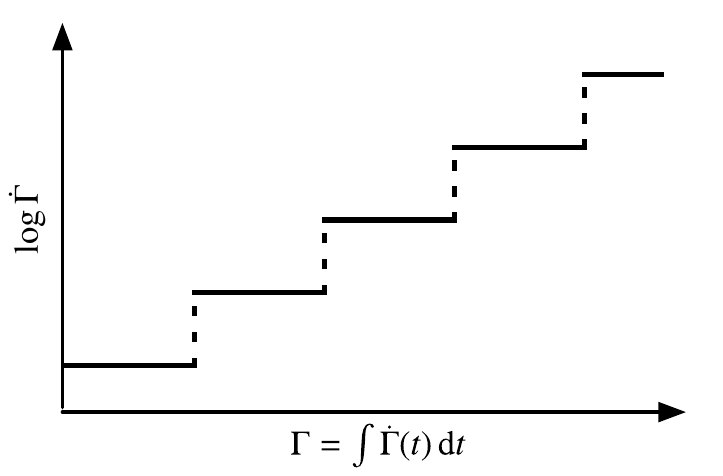}
  \caption{%
    The shear rate $\dot{\Gamma}$ 
    is increased in a stepwise manner.
    The horizontal axis shows the total shear strain.
  }
\label{stepwise-shear-rates}
\end{figure}

In this work,
the focus is placed on
such restructuring and consolidation aspects.
So, the flow is turned on less abruptly.
In order to plot the intermediate states of clusters by shear rates,
the shear rate is increased in a stepwise manner.
The $k$-th shear rate is given by
\begin{equation}
  \dot{\Gamma}_k =  \dot{\Gamma}_{1}
  \bigl(
    \dot{\Gamma}_{\mathrm{max}} / \dot{\Gamma}_{1}
    \bigr)^{(k-1)/(k_{\mathrm{max}}-1)},
\end{equation}
where $\dot{\Gamma}_{\mathrm{1}}$ is the initial shear rate,
$\dot{\Gamma}_{\mathrm{max}}$ the final shear rate
and $k_{\mathrm{max}}$ the number of steps,
and it is kept for the time period $t^{\ast}_k$
resulting in an equivalent total shear strain $\Gamma^{\ast}$,
\textit{i.e.} $t^{\ast}_k = \Gamma^{\ast}/\dot{\Gamma}_{k}$
(see \figref{stepwise-shear-rates}).

\section{Results}

\subsection{Parameters for the simulation}
\label{sec_parameters}
Fractal clusters generated by 
the reaction limited hierarchical cluster-cluster aggregation (CCA)
were used as an initial configuration~\cite{Botet_1984,Jullien_1987}.
The fractal dimension is $d_{\mathrm{f}} \approx 2$.
It is worth noting that 
such generated clusters have no loop structure.
In previous works~\cite{Seto_2011,Seto_2012},
the hydrodynamic behavior of various sizes of the same CCA clusters 
have been examined by assuming rigid structures.
Here, the restructuring behavior of small clusters with $N=64$
was investigated.
%

%%%%%%%%%%%%%%%%%%%%%%%%%%%%%%%%%%%%%%%%%%%%%%%%%%%%%%%%%%%%%%%%%%%%%%%%%%%%%%%%%%%%%%%%%%%%%%%%%%%%

\begin{figure*}[htb]
  \centering
  \includegraphics{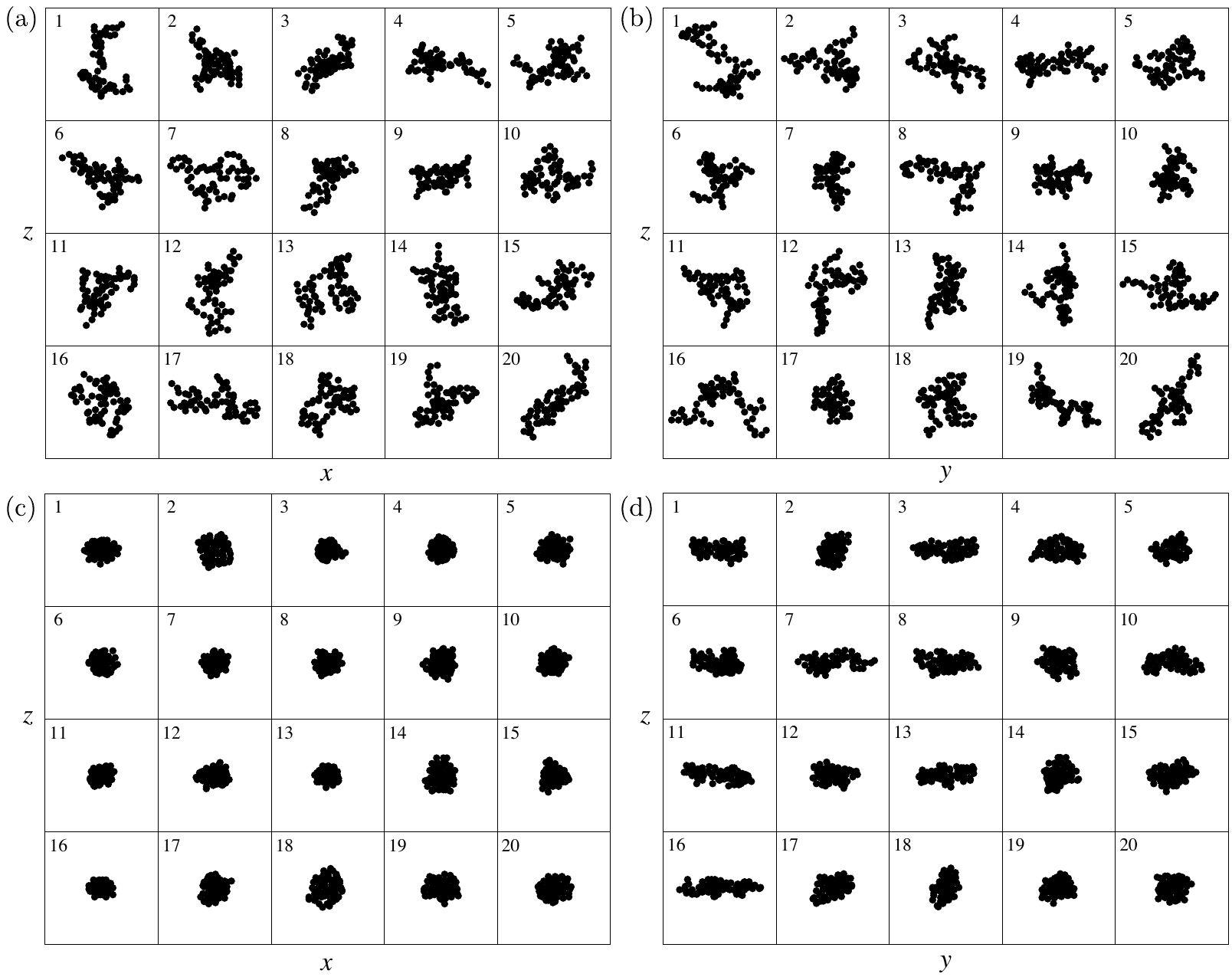}
  \caption{%
    A random selection of the initial clusters (CCA clusters, $N=64$)
    are shown by $x$-$z$ and $y$-$z$ projections (a) and (b),
    and the corresponding compacted clusters after $\dot{\Gamma}=15.9$
    are also shown by $x$-$z$ and $y$-$z$ projections (c) and (d).
  }
  \label{snapshots_cca}
\end{figure*}

For randomly structured clusters,
one needs to evaluate a sufficient number of samples
to study any generalizable behavior,
therefore 50 independent clusters
were simulated under the same conditions.
A random selection of the initial clusters
is shown as projections on $x$-$z$ and $x$-$y$ planes 
in \figref{snapshots_cca} (a) and (b).

%%%%%%%%%%%%%%%%%%%%%%%%%%%%%%%%%%%%%%%%%%%%%%%%%%%%%%%%%%%%%%%%%%%%%%%%%%%%%%%%%%%%%%%%%%%%%%%%%%%%

The required parameters for the contact model (see \secref{sec_contact_model})
are only the ratios between the spring constants of different modes 
and the critical moment.
For the spring constants,
the same value was set for the bending and torsional modes,
and 10 times larger for normal and sliding modes:
\begin{equation}
  k_{\mathrm{T}} = k_{\mathrm{B}},
  \quad
  k_{\mathrm{N}} = k_{\mathrm{S}}  = 10 k_{\mathrm{B}}.  
\end{equation}
The critical moment $M_{\mathrm{c}}$ for 
bending and torsional springs in \eqref{simplified_destruction_functions}
was set to the value which gives 1\% of the particle's radius
for the critical displacements.

%%%%%%%%%%%%%%%%%%%%%%%%%%%%%%%%%%%%%%%%%%%%%%%%%%%%%%%%%%%%%%%%%%%%%%%%%%%%%%%%%%%%%%%%%%%%%%%%%%%%

The used parameters for the imposed shear flows
(see \secref{sec_stepwise_shear})
are presented by \tabref{flow_parameters}.
In order to distinguish the hydrodynamic effect with Stokesian dynamics (SD), 
the free-draining approximation (FDA) was also used as the reference.
For both methods,
the ranges of the shear-rate changes were chosen 
to see the rigid-body rotation regime with the lower shear rates
and the sufficient compaction with the higher shear rates.

%%%%%%%%%%%%%%%%%%%%%%%%%%%%%%%%%%%%%%%%%%%%%%%%%%%%%%%%%%%%%%%%%%%%%%%%%%%%%%%%%%%%%%%%%%%%%%%%%%%%

\begin{figure}[htb]
 \centering
\includegraphics{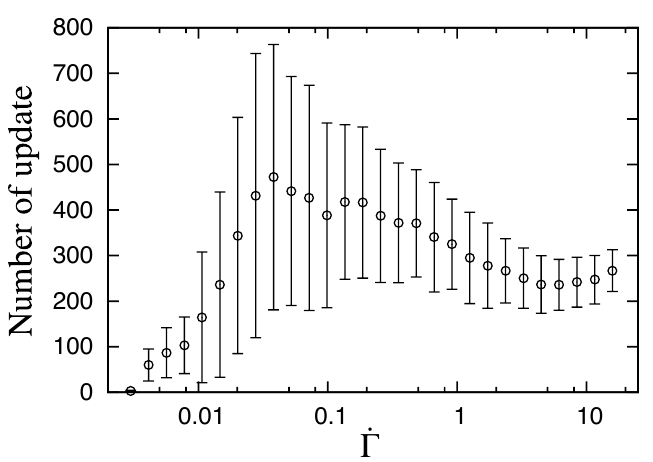}
\caption{
  The mobility matrix is updated when 
  the cluster deformation exceeds the threshold.
  The average numbers of the updates
  during one cluster rotation are shown. 
  The error bars exhibit the standard deviations
  over 50 independent simulations.
}
\label{fig_update_num}

\end{figure}

For reusing the mobility matrix for deformed clusters
(see \secref{sec_optimization}),
the threshold $D_{\mathrm{max}} = 0.01a^2$ was given,
which is small enough to evaluate 
the drag forces in acceptable precision for our purpose.
The actual update numbers of the mobility matrix 
during one cluster rotation are given in \figref{fig_update_num}.
These numbers are much less than
the time steps to integrate the equations of motion,
but 
these updates sufficiently reflect 
the long-range hydrodynamic interactions acting on deforming clusters.

\begin{table}[tbh]
  \caption{
    Parameters of the imposed flows.
  }
  \label{flow_parameters}
  \newcolumntype{C}{>{\centering\arraybackslash}X}
  \newcolumntype{R}{>{\raggedright\arraybackslash}X}
  \newcolumntype{L}{>{\raggedleft\arraybackslash}X}
  \begin{tabularx}{\columnwidth}{lCCC}
    \hline 
    & Symbol & SD & FDA \\
    \hline  
    Initial shear rate  &$\dot{\Gamma}_1$ & 0.003 & 0.001 \\
    Final shear rate  &$\dot{\Gamma}_{\mathrm{max}}$ & 15.9  & 10\\
    Number of steps & $ k_{\mathrm{max}}$ & 28  & 30\\
      Total shear strain & 
      \multirow{2}{*}{$\Gamma^{\ast}$} & \multirow{2}{*}{20}  & \multirow{2}{*}{20} \\
      ~ on time interval &  \\
    \hline
  \end{tabularx} 
\end{table}

\subsection{Compaction}
\label{sec_compaction}
\begin{figure}[hbt]
\centering
\includegraphics{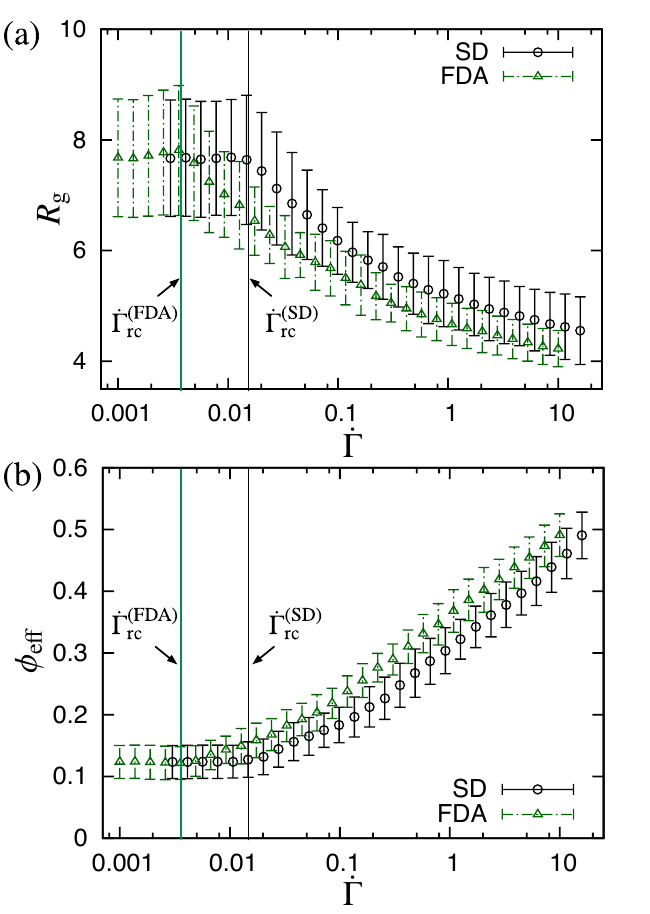}
\caption{
  The compaction behavior is seen by
  the shear-rate dependence of the radius of gyration $R_{\mathrm{g}}$ (a),
  and the effective volume fraction $\phi_{\mathrm{eff}}$ (b).
  The final values at the each shear-rate step were sampled,
  and the averages and standard deviations were taken over 50 independent simulations.
  The results with SD and FDA
  are shown by circles $(\bigcirc)$
  and triangles ($\bigtriangleup$), respectively.
}
\label{fig_compaction}
\end{figure}

First, the relation between the compaction and the flow strength was considered.
The radius of gyration
\begin{equation}
  R_{\mathrm{g}}^2 \equiv 
  \frac{1}{N} \sum_{i=1}^{N} (\vec{r}^{(i)} - \vec{r}_0)^2,
\end{equation}
where $ \vec{r}_0 $ is the center of mass of the cluster,
approximately represents the hydrodynamic radius of the fractal clusters%
~\cite{Wessel_1992,Lattuada_2003a,Seto_2011}.
Indeed, the radius of gyration has been commonly used to quantify the size of 
random structured colloidal aggregates.
\figref{fig_compaction} (a) shows the shear-rate dependence of the radius of gyration,
where final values at each shear-rate step were sampled.
The averages and standard deviations were taken over 50 independent simulations.
However, this quantity is not optimal to address the compaction behavior
because results for compacted clusters depend on their shapes.

%%%%%%%%%%%%%%%%%%%%%%%%%%%%%%%%%%%%%%%%%%%%%%%%%%%%%%%%%%%%%%%%%%%%%%%%%%%%%%%%%%%%%%%%%%%%%%%%%%%%

The volume fraction is an alternative to quantify the compaction 
as it takes into account cluster shapes.
As seen in \figref{snapshots_cca} (c) and (d),
some of compacted clusters exhibit elongated shapes.
Though the definition of volume fraction is not simple for isolated clusters,
a rough estimation was used here.
An arbitrary shaped cluster can be translated
into an ellipsoid having the equivalent moments-of-inertia.
We take the ratio between the total volume of particles
and the volume of the equivalent ellipsoid
as the effective volume fraction $\phi_{\mathrm{eff}}$
(see \appendixname~\ref{equivalent_ellipsoid}).
The shear-rate dependence of the effective volume fraction
is shown in \figref{fig_compaction} (b).
The standard deviations are reduced in comparison with 
the plot for the radius of gyration.
So, the effective volume fraction is used 
as a measure for the compaction behavior.

%%%%%%%%%%%%%%%%%%%%%%%%%%%%%%%%%%%%%%%%%%%%%%%%%%%%%%%%%%%%%%%%%%%%%%%%%%%%%%%%%%%%%%%%%%%%%%%%%%%%
For lower shear rates,
both $R_{\mathrm{g}}$ and $\phi_{\mathrm{eff}}$ are almost constant,
which indicates the `rigid body rotation' regime described in \secref{sec_stepwise_shear}.
By increasing the shear rate,
the compaction starts at a certain point.
This critical shear rate is denoted by $\dot{\Gamma}_{\mathrm{rc}}$.
For reference, the results by using the free-draining approximation (FDA)
are also plotted by the triangle marks ($\bigtriangleup$) in \figref{fig_compaction}.
The critical shear rate by FDA is much smaller,
because the imposed flows are not disturbed by the cluster in FDA.
Though the cluster size is small ($N=64$),
the ratio of the critical shear rates for the two methods is significant
$\dot{\Gamma}_{\mathrm{rc}}^{\mathrm{(SD)}}/
\dot{\Gamma}_{\mathrm{rc}}^{\mathrm{(FDA)}}\approx 4.2$.

%%%%%%%%%%%%%%%%%%%%%%%%%%%%%%%%%%%%%%%%%%%%%%%%%%%%%%%%%%%%%%%%%%%%%%%%%%%%%%%%%%%%%%%%%%%%%%%%%%%%

Under stepwise increasing shear flows, clusters are monotonously compacted.
The shear-rate dependence of the effective volume fraction
turned out to be rather small
so that the maximum compaction ($\mathrm{d} \phi_{\mathrm{eff}}/\mathrm{d} \dot{\Gamma} \to 0$)
was not observed within the simulated range of the shear rates.
The higher shear rate required for further compaction
may violate the model assumptions,
such as Stokes regime for the hydrodynamics (\secref{sec_method_SD})
and the conditions for the overdamped motion (\secref{sec_overdamped_motion}).
The maximum compaction in the SD simulations
is $\phi_{\mathrm{eff}} \approx 0.49$ with $\dot{\Gamma}^{\mathrm{(SD)}}  \approx 15.9$.
The equivalent compaction by using FDA requires
a shear rate of $\dot{\Gamma}^{\mathrm{(FDA)}} \approx 10.0$.
It is worth noting that
the range of shear rates for this compaction is quite wide,
\textit{i.e.}
$\dot{\Gamma}^{\mathrm{(SD)}} (\phi = 0.49)/ \dot{\Gamma}_{\mathrm{rc}}^{\mathrm{(SD)}}  
\approx 1.1 \times 10^3$.
A discussion about this point 
will be given later (\secref{sec_consolidation}).

\subsection{Shape and orientation during compaction}
\label{sec_shape_orientation}
The simulation results showed some characteristic ways of the clusters' compaction.
In \figref{snapshots_cca},
the initial configurations, (a) and (b),
and the corresponding compacted clusters, (c) and (d), are displayed.
It can be noticed that
the initial clusters have a variety of shapes and orientations,
while the compacted clusters can be roughly classified to two types:
rod-shaped clusters elongating to $y$ axis and round-shaped clusters.
This section deals with
the transitions of the shapes and orientations during the compaction.
%

%%%%%%%%%%%%%%%%%%%%%%%%%%%%%%%%%%%%%%%%%%%%%%%%%%%%%%%%%%%%%%%%%%%%%%%%%%%%%%%%%%%%%%%%%%%%%%%%%%%%

\begin{figure*}
  \includegraphics{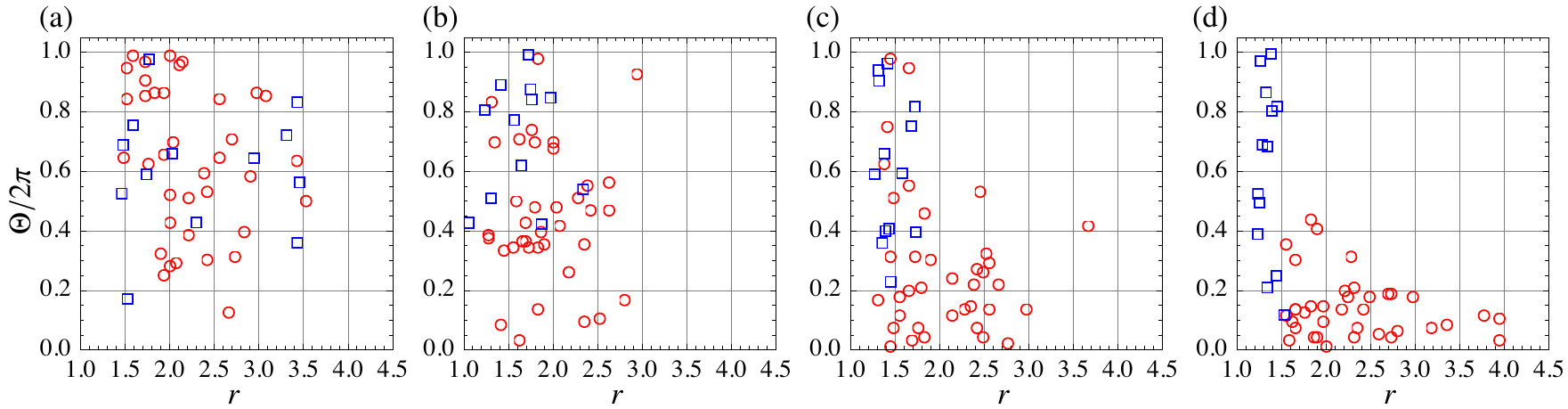}
  \caption{
    The scatter plots of the aspect ratio \textit{vs.} tilting angle ($r$-$\Theta$) 
    representations for the initial configurations (a), 
    for the results at two intermediate shear rates
    $\dot{\Gamma}=0.99\times 10^{-1}$ (b)
    and $\dot{\Gamma}=0.91 $ (c),
    and for the final configuration with $\dot{\Gamma}=1.59\times 10$ (d),
    show the restructuring processes of 50 simulations.
    The circle ($\bigcirc$) and square ($\Box$) plots 
    indicate the groups of rod-shaped and round-shaped, respectively.
    For (b)-(d), the final values at the each shear-rate step were sampled.
  }
  \label{shape_orientation}
\end{figure*}

In order to quantify the slenderness of clusters,
the aspect ratio of the equivalent ellipsoid $r$
is used (see \appendixname~\ref{equivalent_ellipsoid}).
The orientations can be quantified by 
the tilting angle $\Theta$,
which is the angle between 
the principal axis $\vec{n}_1$ of the principal moment of inertia with the smallest value 
and $y$ axis:
\begin{equation}
  \Theta \equiv
  \arccos \bigl(\vec{e}_{y} \cdot \vec{n}_{1}\bigr).  
\end{equation}
Now, the transitions of the shape and orientation of the distinct simulations
can be represented by the aspect ratio \textit{vs.} tilting angle ($r$-$\Theta$) 
scatter plot.
The four stages of the compaction,
\textit{i.e.} the initial (a), two intermediate (b) and (c), 
and final (d) configurations,
are shown in \figref{shape_orientation}.

%%%%%%%%%%%%%%%%%%%%%%%%%%%%%%%%%%%%%%%%%%%%%%%%%%%%%%%%%%%%%%%%%%%%%%%%%%%%%%%%%%%%%%%%%%%%%%%%%%%%

As mentioned above, 
the clusters seem to be distinguished by two types after the compaction.
In order to follow the formation processes,
they are separated into two groups by introducing a selection criterion.
The clusters having aspect ratios $r > 1.56$
at the results with $\dot{\Gamma} = 15.9$
are classified as rod-shaped clusters and the others as round-shaped clusters.
The threshold value $r = 1.56$ was chosen
by considering the distribution of the orientations,
\textit{i.e.} clusters having smaller aspect ratio than this value 
no longer show an orientation tendency.
For the simulated clusters, 74\% of them are classified as rod-shaped.
In \figref{shape_orientation},
the circle ($\bigcirc$) and square ($\Box$)
plots indicate the clusters ending up as the rod-shaped and round-shaped clusters,
respectively.
%

%%%%%%%%%%%%%%%%%%%%%%%%%%%%%%%%%%%%%%%%%%%%%%%%%%%%%%%%%%%%%%%%%%%%%%%%%%%%%%%%%%%%%%%%%%%%%%%%%%%%
The representation for the initial clusters [\figref{shape_orientation} (a)]
shows the following:
The CCA clusters originally have anisotropic structures,
so that the aspect ratios are distributed between 1.5 and 3.5.
Their orientations, on the other hand, are uniformly distributed.
In this representation, 
the two groups appears to be randomly mixed.
Thus, at the moment 
we cannot predict which mode
of compaction from the initial configuration.
%

%%%%%%%%%%%%%%%%%%%%%%%%%%%%%%%%%%%%%%%%%%%%%%%%%%%%%%%%%%%%%%%%%%%%%%%%%%%%%%%%%%%%%%%%%%%%%%%%%%%%

\begin{figure*}
  \centering
  \includegraphics{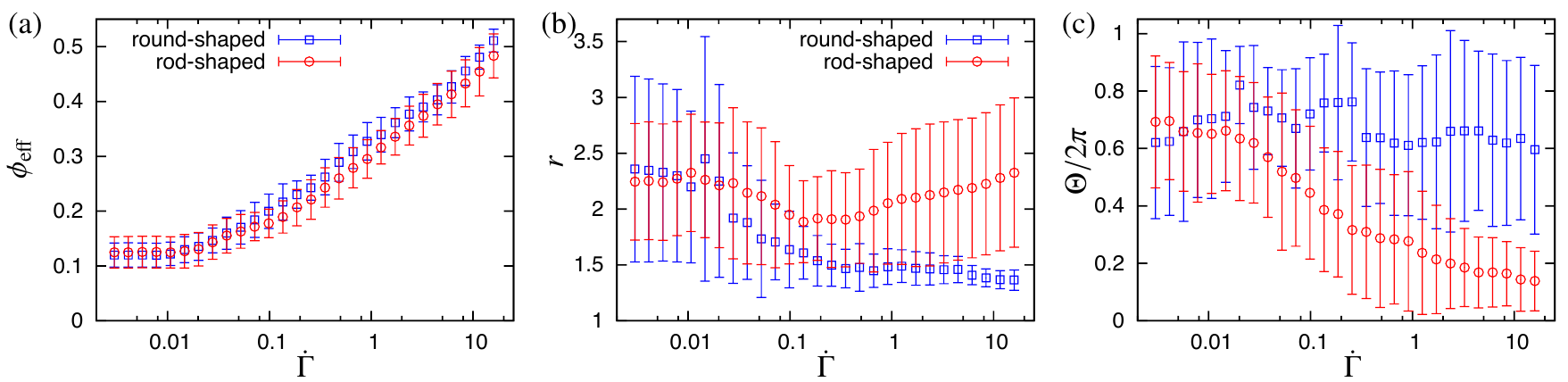}
  \caption{
    The shear-rate dependence
    of the effective volume fraction $\phi_{\mathrm{eff}}$ (a),
    aspect ratio $r$ (b), and tilting angle $\Theta$ (c),
    are shown by the two groups:
    rod-shaped and round-shaped, respectively.
    The error bars indicate the standard deviations 
    over the members of the respective groups.
    The final values at the each shear-rate step were sampled.
  }
  \label{ave_std_transitions}
\end{figure*}

The compaction processes can be followed
by seeing the translations of the plots
from (a) to (d) in \figref{shape_orientation}.
Besides, 
in order to see the shear-rate dependence,
the averages and standard deviations 
of the respective quantities by each group
are shown in \figref{ave_std_transitions}.
For the effective volume fractions $\phi_{\mathrm{eff}}$,
the difference between two groups is not significant%
~[\figref{ave_std_transitions}~(a)].
For the clusters ending up as rod-shaped clusters,
the distribution of the aspect ratio $r$ is slightly compressed 
to smaller values at the beginning,
and it is shifted to larger values after $\dot{\Gamma} \approx 0.14$~%
[\figref{ave_std_transitions} (b)].
The distribution of the orientation shows a more systematic change,
\textit{i.e.}
the reorientation seems to correlate with the compaction~%
[\figref{ave_std_transitions} (c)].
For the clusters ending up as the round-shaped clusters,
the aspect ratio is decreased for $\dot{\Gamma} < 0.2$,
and stays constant after that [\figref{ave_std_transitions} (b)].

%%%%%%%%%%%%%%%%%%%%%%%%%%%%%%%%%%%%%%%%%%%%%%%%%%%%%%%%%%%%%%%%%%%%%%%%%%%%%%%%%%%%%%%%%%%%%%%%%%%%

\begin{figure}[htb]
  \centering
  \includegraphics{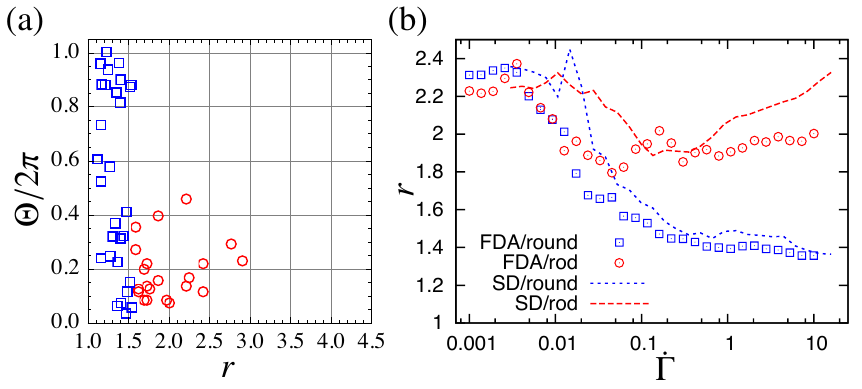}
  \caption{
    (a) The aspect ratio \textit{vs.} tilting angle ($r$-$\Theta$) representations
    for the compacted clusters ($\dot{\Gamma}=10$) by using FDA are shown.
    The circle ($\bigcirc$) and square ($\Box$) plots indicate 
    the clusters ending up as the rod-shaped and round-shaped clusters, respectively.
    (b) The aspect ratios $r$ are averaged over the members of the respective groups,
    which show the shear-rate dependence.
    The results by using SD are also shown by the dashed lines to compare with FDA.
    \label{FDA}
  }
\end{figure}

For reference, the results with FDA are also shown in \figref{FDA}.
The observed tendency seen in \figref{shape_orientation} (d)
is less clear in the equivalent compaction with FDA [\figref{FDA} (a)].
For the same selection criteria,
42\% of the clusters are classified as the rod-shaped group.
The averages of the aspect ratio are taken over each group,
and the shear-rate dependence is presented in \figref{FDA}~(b).
For the result of FDA, 
the aspect-ratio of the rod-shaped clusters
does not increase as in the SD simulations (shown by the dashed line).
Instead it remains almost constant.
Thus, by comparing simulations of SD and FDA,
it turns out that the features seen with SD are less pronounced with FDA.

%%%%%%%%%%%%%%%%%%%%%%%%%%%%%%%%%%%%%%%%%%%%%%%%%%%%%%%%%%%%%%%%%%%%%%%%%%%%%%%%%%%%%%%%%%%%%%%%%%%%

%
\section{Discussion}
\label{sec_discussion}

\subsection{Consolidation}
\label{sec_consolidation}

As seen in \ref{sec_compaction},
the shear rates for the compaction of clusters 
range over multiple  orders of magnitude,
\textit{i.e.} initial fractal clusters are fragile
while compacted clusters are more and more robust to imposed flows.
In our simulation with Stokesian dynamics (SD), 
the highest shear rate (resulting in $\phi_{\mathrm{eff}}\approx 0.49$)
is about $10^3$ times larger than 
the critical shear rate $\Gamma_{\mathrm{rc}}$ where $\phi_{\mathrm{eff}}\approx 0.12$.
This may be explained by 
the following two non-linear effects:
\begin{enumerate}
  \item[(i)] 
  The smaller the hydrodynamic radius is,
  the weaker the hydrodynamic stress acting on the cluster becomes
  \item[(ii)] 
  The higher the number of newly generated loops within the cluster,
  the more the cluster resists restructuring.
\end{enumerate}
%
%%%%%%%%%%%%%%%%%%%%%%%%%%%%%%%%%%%%%%%%%%%%%%%%%%%%%%%%%%%%%%%%%%%%%%%%%%%%%%%%%%%%%%%%%%%%%%%%%%%%

\begin{figure}[htbp]
\centering
\includegraphics{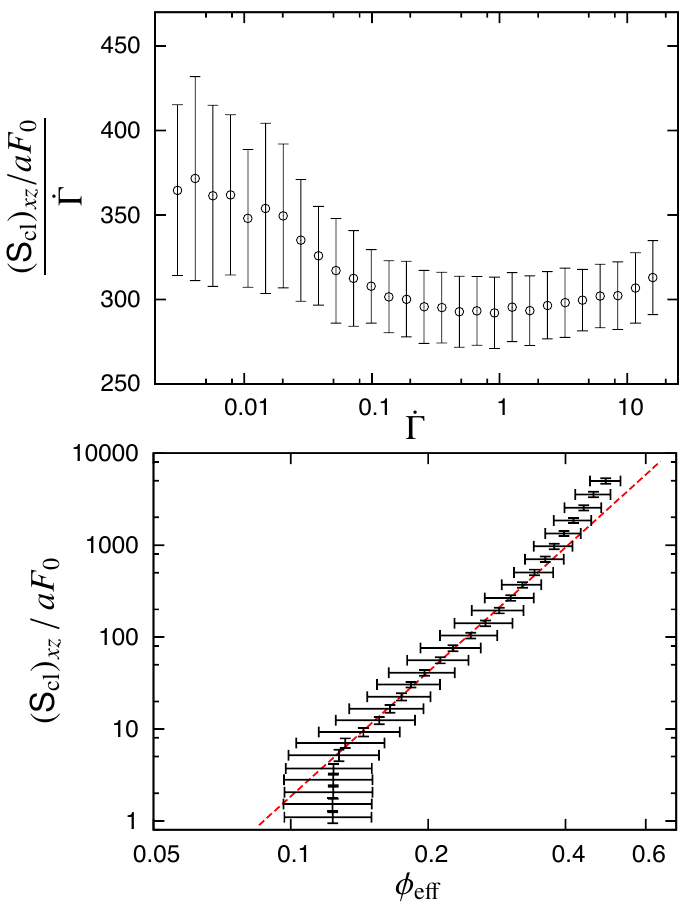}
\caption{
  (a) 
  The shear-rate dependence of 
  the ratio between the stresslet $(\tens{S}_{\mathrm{cl}})_{xz}$
  and the shear rate $\dot{\Gamma}$ are shown.
  (b) 
  The volume-fraction dependence
  of the stresslets acting on clusters are plotted.
  For both the plots,
  the final values at the each shear-rate step were sampled,
  and the error bars indicate the standard deviations.
}
\label{stresslet}
\end{figure}

In order to see effect (i),
the shear-rate dependence of the stresslet acting on a cluster was  evaluated.
The stresslet acting on a cluster is composed as follows~\cite{Harshe_2010,Seto_2011}:
\begin{equation}
  \tens{S}_{\mathrm{cl}} = 
  \sum_{i}
  \bigg\{
    \tens{S}^{(i)}_{\mathrm{H}} +  
    \frac{
      \vec{l}^{(i)}  \otimes \vec{F}^{(i)}
      +   (\vec{l}^{(i)}  \otimes \vec{F}^{(i)})^{\mathrm{T}}
    }{2} 
    - \frac{
	\vec{l}^{(i)} \cdot \vec{F}^{(i)}}{3}
    \tens{I}
    \bigg\},  
\end{equation}
where $\vec{l}^{(i)} \equiv \vec{r}^{(i)}-\vec{r}_0$.
The stresslet $\tens{S}^{(i)}_{\mathrm{H}}$ for individual particles $i$
has already been calculated in \eqref{mobility_form}.
This stresslet $\tens{S}_{\mathrm{cl}}$
indicates the contribution of a single cluster
to the bulk stress of a sheared suspension.
For dilute suspensions of rigid spheres,
the stresslet is proportional to the shear rate: 
$(\tens{S}_{\mathrm{sph}})_{xz} =(20/3)\pi \eta_0 a^3 \dot{\gamma}$.
This is why 
the effect of restructuring on the hydrodynamic stress
appears in the ratio between $(\tens{S_{\mathrm{cl}}})_{xz}$ and $\dot{\Gamma}$.
With the shrinkage of clusters, the efficiency is decreased [\figref{stresslet} (a)].
However, this decrease is not large enough to explain 
the wideness of the shear-rate range.
%
%%%%%%%%%%%%%%%%%%%%%%%%%%%%%%%%%%%%%%%%%%%%%%%%%%%%%%%%%%%%%%%%%%%%%%%%%%%%%%%%%%%%%%%%%%%%%%%%%%%%

The volume-fraction dependence of the stresslet
was also plotted to see effect (ii) [\figref{stresslet} (b)].
If the total strain on time interval $\Gamma^{\ast}$ is infinitely large,
the compaction of cluster at each shear rate 
is expected to be settled after some restructuring,
and the further compaction requires a higher hydrodynamic stress.
This is why
this plot approximately represents 
%the hydrodynamic stress to proceed the compaction of clusters
the compactive strength (yield stress)
as a function of the volume fraction.
Though the hydrodynamic stress given by the stresslet
is not the same quantity as the mechanical stress,
one may notice some similarity 
with volume-fraction dependent compressive yield stress $P_{\mathrm{y}}(\phi)$ 
of space-filling colloidal aggregate networks~\cite{Buscall_1987,Buscall_1988}.
The power-law behavior seen in the intermediate range 
of \figref{stresslet} (b)
shows the large exponent $4.5$,
which was evaluated by fitting the averaged data 
within $0.13 < \phi_{\mathrm{eff}} < 0.34$.
According to the cited work~\cite{Buscall_1988},
large exponents of power-law relations were explained 
as a result of network densification due to irreversible restructuring.
If the hydrodynamic stresses induce deformation of clusters,
the same consolidation effect is
expected by the rule of new bond generations given in \secref{method_new_bond}.
Thus, it was confirmed that the effect (ii) is responsible 
to explain the wideness of the shear-rate range for the compaction.

\subsection{Reorientation and anisotropic compaction}
\label{sec_reorientation}

Rod-shaped clusters orienting to the rotational axis
were observed in the simulations (see \ref{sec_shape_orientation}).
Since the Brownian motion was not taken into account,
the pioneer works by Jeffery on spheroids in shear flows
may be referred to~\cite{Jeffery_1922,Happel_1965,Kim_1991}.
Spheroids can be considered as one of 
the simplest object representing elongated shapes.
Jeffery analytically showed that, for the dilute limit,
they should be in the periodic orbits
and have no tendency to set their axis in any particular direction
under a simple shear flow.
Besides, he expected that,
since the dissipation of energy depends on their orbits,
they would tend to adopt the orbital motion of the least dissipation of energy 
with additional elements in real suspensions.

%%%%%%%%%%%%%%%%%%%%%%%%%%%%%%%%%%%%%%%%%%%%%%%%%%%%%%%%%%%%%%%%%%%%%%%%%%%%%%%%%%%%%%%%%%%%%%%%%%%%

\begin{figure}[htb]
  \centering
  \includegraphics{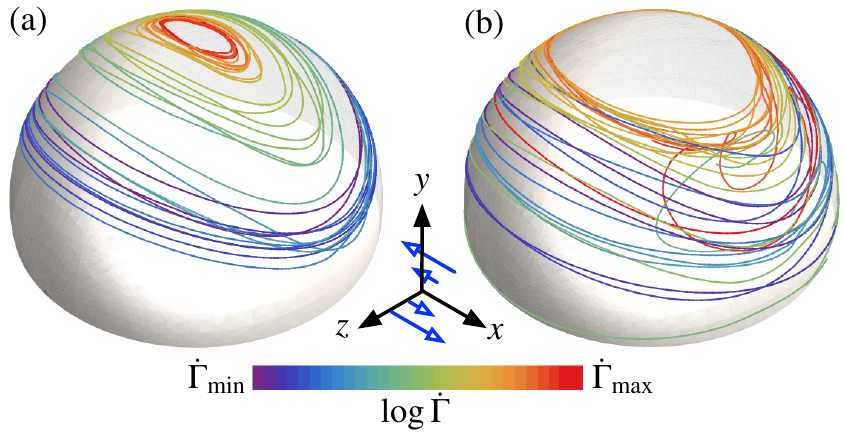}
  \caption{
    (Color online)
    The trajectories of the principal axis 
    for the smallest principal moment of inertia $\vec{n}_1$ 
    are plotted on the unit sphere.
    The two examples are shown:
    (a) the rod-shaped compaction,
    which starts from $(r,\Theta/2\pi)=(2.4,0.59)$ and ends to $(2.7,0.18)$,
    and (b) the round-shaped compaction from $(2.3,0.42)$ to $(1.4,0.67)$.
    The north pole shows the rotational axis ($y$-axis).
    The color scale of the trajectories 
    represents the changes of the shear rates.
  }
  \label{trajectories}
\end{figure}

The trajectories of the principal axis for the smallest principal 
moment of inertia $\vec{n}_1$ are plotted in \figref{trajectories}
for two typical cases 
ending up as the rod-shaped (a) and round-shaped (b) clusters.
Each of them shows the trajectory of 
one simulation with increasing shear rates.
The color scale of the trajectories 
represents the changes of the shear rates.
Though the orbits are not closed,
one can find some similarity with the Jeffery orbits
in the short time behavior~(c.f. Figure 5.5 of ref.~\citep{Kim_1991}).
For the rod-shaped compaction (a), 
the orbit tends to converge in a narrow circuit around the north pole.
It can be noticed that in the round-shaped compaction (b),
the orbit is easier to be affected by the irregular structure of the cluster,
in particular when it crosses the $x$-$y$ plane. 
Thus, if the orientation of clusters is changeable,
the uniform compaction can be expected.

%%%%%%%%%%%%%%%%%%%%%%%%%%%%%%%%%%%%%%%%%%%%%%%%%%%%%%%%%%%%%%%%%%%%%%%%%%%%%%%%%%%%%%%%%%%%%%%%%%%%

Though the randomness of the cluster may lead to uniform compaction,
the formations of the rod-shaped clusters 
and their reorientations are not explained yet.
Another view point is the anisotropic compaction in shear flows.
For clusters being restructured,
the principal axis $\vec{n}_1$ of the cluster is no longer fixed to the cluster,
but depends on the structure at each instant.
If the compaction is anisotropic,
it may look as the reorientation of the principal axis.
In a shear flow,
the drag forces acting on particles within rotating clusters 
increase with the distance from the rotational axis~\citep{Seto_2011}.
So, the displacements of particles
depend on their positions as well.
Since the compaction is caused by generations of cohesive bonds,
one can expect that anisotropic compaction reduces distances
of particles from the rotational axis.
Thus, as long as the rotational axis is unchanged,
the clusters tend to be compacted to elongated shapes.

\subsection{Hydrodynamic effect}
\label{sec_hydrodynamic_effect}

In order to pronounce the hydrodynamic effect, 
the free-draining approximation (FDA) was compared with SD.
First, the hydrodynamic effect is clearly seen
in the hydrodynamic stress for tenuous clusters,
\textit{i.e.} the critical shear rate with SD was much larger than the one with FDA:
$\dot{\Gamma}_{\mathrm{rc}}^{\mathrm{(SD)}}/\dot{\Gamma}_{\mathrm{rc}}^{\mathrm{(FDA)}} 
\approx 4.2$~(see \secref{sec_compaction}).
In low Reynolds number flows,
the disturbance of flows decays proportional 
to the inverse of the distance~\cite{Happel_1965,Kim_1991},
which results in the reduction of the drag forces acting on particles 
within isolated clusters~\cite{Seto_2011}.
Second,
the difference between the two methods
was also confirmed in the shape and orientation tendencies 
by the compaction (\ref{sec_shape_orientation}.)
The spatial distribution of the drag force within clusters 
has the same symmetry for FDA and SD~\citep{Seto_2011}.
This is why some qualitative explanations for the formation of rod-shaped clusters
are expected to be applicable for the simulation with not only SD but also FDA.
However, as seen in \figref{FDA},
the result with FDA did not show clear tendency to form rod-shaped clusters.
This result suggests that 
the hydrodynamic effect works as a kind of positive feedback
for the anisotropic compaction in shear flows.
\section{Conclusion}
\label{sec_conclusion}

Restructuring of colloidal aggregates in shear flows
has been investigated by coupling an interparticle contact model 
with Stokesian dynamics.
We have introduced a method 
to reduce calculation cost for near-rigid behavior of aggregates in shear flows.
That method has realized fluid-particle coupling simulations
for a significant time period.
The simulations with the stepwise increase of shear rate
have demonstrated the reinforcement of clusters due to irreversible compaction
with increase of the hydrodynamic stress.
We expect that the observed consolidation behavior 
induced by less-abrupt-application of flows is rather general.
The introduced aspect ratio \textit{vs.}~orientation representation
has also clarified 
the two types of compaction behaviors
ending up in rod-shaped clusters orienting to the rotational axis
and to round-shaped clusters.
This anisotropic compaction can be a sort of hydrodynamic effect,
\textit{i.e.} the enhancement of the tendency was
seen in the comparison with the free-draining approximation.
Thus,
the simulations with selected parameters for the contact model
showed the characteristic behaviors 
of colloidal aggregates under flow conditions.

\begin{acknowledgement}
The authors would like 
to thank Dr. Martine Meireles
and Prof. Bernard Cabane
for many valuable discussions on the contact model
and Dr.~K. Ichiki
for providing the simulator of Stokesian dynamics ``Ryuon'' and instructive advice,
and Vincent Bürger for assistance with the manuscript. 
We acknowledge the financial support of the German Science
Foundation (DFG priority program SPP 1273).
\end{acknowledgement}
\appendix
\section{Descriptions by the equivalent ellipsoid}
\label{equivalent_ellipsoid}

In order to quantify the shape and orientation 
of random-structured clusters,
we translate them to equivalent ellipsoids
having the same principal moments of inertia~\citep{Harshe_2010}.
The principal moments of inertia $I_1$, $I_2$, $I_3$ ($I_1 \leq I_2 \leq I_3$)
are obtained by diagonalization of the moment of inertia tensor.
By using them,
the lengths of the semi-principal axes of the equivalent ellipsoid
($a \geq b \geq c $) are given as follows:
\begin{gather}
    a = \sqrt{\frac{5}{2} \frac{I_2 + I_3 - I_1}{N}}, \quad  
    b = \sqrt{\frac{5}{2} \frac{I_3 + I_1 - I_2}{N}}, \notag \\
    c = \sqrt{\frac{5}{2} \frac{I_1 + I_2 - I_3}{N}}.    
\end{gather}
The anisotropic shape of clusters
is described by the aspect ratio $r$ of the equivalent ellipsoid:
\begin{equation}
r \equiv \frac{2a}{b+c},
\end{equation}
In order to estimate the compaction of clusters,
the effective volume fraction $ \phi_{\mathrm{eff}} $
is introduced as the ratio between
the total volume of particles $V_{\mathrm{p}}=(4\pi/3)N$
and the volume of the ellipsoid $V_{\mathrm{e}} = (4\pi/3) a b c$,
that is
\begin{equation}
  \phi_{\mathrm{eff}} 
 \equiv \frac{N}{a b c}.
\end{equation}
%

%\bibliography{/Users/seto/Dropbox/Papers/rse}
\end{document}